\documentclass[11pt]{article}

\usepackage{a4wide}

\usepackage{color}
\usepackage{graphicx}
\usepackage{latexsym}
\usepackage[algoruled,vlined,algosection]{algorithm2e}
\SetKwInOut{Input}{Input}
\SetKwInOut{Output}{Output}

\newtheorem{thm}{Theorem}
\newtheorem{lemma}[thm]{Lemma}

\newtheorem{propo}[thm]{Proposition}
\newtheorem{cor}[thm]{Corollary}

\newcommand{\klob}{{\sc $k$-Leaf Out-Branching}}
\newcommand{\klot}{{\sc $k$-Leaf Out-Tree}}

\newcommand{\lot}{\ell}
\newcommand{\lob}{\ell_s}

\newcommand{\BR}{\mbox{\sc Branch}}

\newcommand{\LD}{\mbox{\sc Dist}_L}

\newcommand{\LL}{\mbox{\sc Leaf}}
\newcommand{\BO}{\mbox{\sc BrSucc}}
\newcommand{\Hd}{\mbox{\sc Head}}
\newcommand{\Back}{\mbox{\sc Back}}

\newcommand{\bs}{\backslash}

\newcommand{\proof}{{\bf Proof: }}
\newcommand{\QED}{\mbox{}\hspace*{\fill}{$\Box$}\medskip}

\title{Tight Bounds and Faster Algorithms for Directed Max-Leaf Problems}

\author{Paul Bonsma\footnote{Supported by the Graduate School ``Methods for Discrete Structures'' in Berlin, DFG grant GRK 1408.}\\
\small Technische Universit\"{a}t Berlin,\\
\small Institut f\"{u}r Mathematik, Sekr. MA 5-1,\\ 
\small Stra\ss{}e des 17. Juni 136, 10623 Berlin, Germany\\
\small bonsma@math.tu-berlin.de\\
\\
Frederic Dorn\\
\small Humboldt-Universit\"{a}t zu Berlin\\
\small Institut f\"{u}r Informatik\\
\small Unter den Linden 6, 10099 Berlin, Germany\\
\small dorn@informatik.hu-berlin.de
}

\date{\today}

\begin{document}

\maketitle

\begin{abstract}
An out-tree $T$ of a directed graph $D$ is a rooted tree subgraph with all arcs directed outwards from the root. An out-branching is a spanning out-tree. By $\lot(D)$ and $\lob(D)$ we denote the maximum number of leaves over all out-trees and out-branchings of $D$, respectively.

We give fixed parameter tractable algorithms for deciding whether $\lob(D)\geq k$ and whether $\lot(D)\geq k$ for a digraph $D$ on $n$ vertices, both with time complexity $2^{O(k\log k)} \cdot n^{O(1)}$. This improves on previous algorithms with complexity $2^{O(k^3\log k)} \cdot n^{O(1)}$ and $2^{O(k\log^2 k)} \cdot n^{O(1)}$, respectively.

To obtain the complexity bound in the case of out-branchings, we prove that when all arcs of $D$ are part of at least one out-branching, $\lob(D)\geq \lot(D)/3$. The second bound we prove in this paper states that for strongly connected digraphs $D$ with minimum in-degree 3, $\lob(D)\geq \Theta(\sqrt{n})$, where previously $\lob(D)\geq \Theta(\sqrt[3]{n})$ was the best known bound. 
This bound is tight, and also holds for the larger class of digraphs with minimum in-degree 3 in which every arc is part of at least one out-branching.
\end{abstract}

\section{Introduction}
\label{sec:intro}

Many important graph problems are well-studied on undirected graphs unlike their generalizations to directed graphs. 
One reason may be that despite their practical significance, it is generally harder to obtain similar results for directed graphs.
{\sc Max-Leaf Spanning Tree} is such a problem that has received a lot of study, both algorithmically and combinatorial.
This optimization problem is defined as follows: given an undirected graph, find a spanning tree with maximum number of leaves. In the decision version of this problem, in addition an integer $k$ is given, and the question is whether a spanning tree with at least $k$ leaves exists ({\sc $k$-Leaf Spanning Tree}).
These problems are motivated by many practical and theoretical applications (see for instance~\cite{GK98,WuChao03}), though in some cases, the applications call for a directed generalization~\cite{ICALP08hopefully},
which is what we study in this paper.

For directed graphs or {\em digraphs} we use notions that are defined for undirected graphs, such as paths, trees, connectedness and vertex neighborhoods. These are defined as expected, where arc directions are ignored.
An \emph{out-tree} of a digraph $D$ is a tree subgraph where every vertex has in-degree 1 except for one, the {\em root}, which has in-degree 0. 
An \emph{out-branching} is a spanning out-tree.
A {\em leaf} is a vertex with out-degree 0. In the directed generalization of the problem, one asks for an out-branching with maximum number of leaves. This problem is called {\sc Max-Leaf Out-Branching}. By $\lot(D)$ and $\lob(D)$ we denote the maximum number of leaves over all out-trees and out-branchings of $D$ respectively (when considering $\lob(D)$ we assume that $D$ has at least one out-branching).
Clearly $\lot(D)\geq \lob(D)$ holds, but in contrast to undirected graphs, we do not always have equality here. In fact the ratio $\lot(D)/\lob(D)$ can be arbitrarily large.
 Therefore, on digraphs, the problem of finding an out-tree with maximum number of leaves ({\sc Max-Leaf Out-Tree}) is of independent interest. The corresponding decision problems where the question is asked whether $\lob(D)\geq k$ or whether $\lot(D)\geq k$ are called {\sc $k$-Leaf Out-Branching} and {\sc $k$-Leaf Out-Tree}, respectively.
The related problem of finding out-branchings with {\em minimum} number of leaves has also been considered recently~\cite{Razgon08}. 
 In the first part of this paper we are concerned with algorithmic questions, and in the second part we study the combinatorial question of finding lower bounds for $\lot(D)$ and $\lob(D)$. Throughout this section $n$ denotes the number of vertices of the graph under consideration.

The $NP$-hardness of all problems above follows from the $NP$-completeness of {\sc $k$-Leaf Spanning Tree}.
Whereas for the undirected problem, {\sc Max-Leaf Spanning Tree}, a $2$-approximation is known~\cite{Solis-Oba98}, the best known approximation result for {\sc Max-Leaf Out-Branching} is a very recent algorithm with ratio $O(\sqrt{n})$~\cite{DrescherV07}.
In the algorithmic part of this work, we are interested in \emph{fixed parameter tractable (FPT)} algorithms for the decision problems. We choose the desired number of leaves $k$ as the parameter. Then an algorithm is an {\em FPT algorithm} if its time complexity is bounded by a function of the form $f(k) \cdot n^{O(1)}$, where the {\em parameter function} $f$ may be any computable function only depending on $k$. FPT algorithms are well-studied and classified. 
The book of Downey and Fellows~\cite{DowneyF99} provides an introduction into
 parameterized complexity. See the books of Flum and Grohe~\cite{FlumGrohebook} and Niedermeier~\cite{Niedermeierbook06} for more recent introductions into parameterized complexity.
The main indicator of the practicality of FPT algorithms is the growth rate of the parameter function, and which is one important reason to design FPT algorithms with small parameter function. For the undirected problem {\sc $k$-Leaf Spanning Tree} many improvements have been made in this area (see e.g.~\cite{FellowsMRS00,BonsmaBW03}), which has also has been a large stimulus for research on related combinatorial questions. The current fastest FPT algorithm has a running time of $O^*(6.75^k)+O(m)$, with $m$ being the number of edges~\cite{bonsmazickfeld}.

Considering the directed versions of the problem, 
Alon et al~\cite{AlonFGKS07} were the first to give an FPT algorithm for {\sc $k$-Leaf Out-Tree}, which had a running time $2^{O(k^2 \log k)} \cdot n^{O(1)}$. In~\cite{AlonFGKS07II} they improved this to $2^{O(k \log^2 k)} \cdot n^{O(1)}$.
They also observed that for digraph classes where out-trees with $k$ leaves can always be extended to out-branchings with $k$ leaves, this solves {\sc $k$-Leaf Out-Branching} with the same time complexity, and that this property holds for the important classes of strongly connected digraphs and acyclic digraphs. For acyclic 
digraphs, they also gave  a specialized algorithm for {\sc $k$-Leaf Out-Branching} with a complexity of $2^{O(k \log k)} \cdot n^{O(1)}$~\cite{AlonFGKS07II}. 
Only very recently, the question whether an FPT algorithm exists for \klob\ for all digraphs 
has been resolved, by giving an algorithm with complexity $2^{O(k^3 \log k)}\cdot n^{O(1)}$~\cite{ICALP08hopefully}.

\medskip

{\em In this paper we present FPT algorithms for both \klot\ and \klob\ with parameter function $2^{O(k\log k)}$.} This 
improves the complexity of all FPT algorithms for digraphs mentioned above, except for the algorithm for acyclic 
digraphs, which has the same complexity.

 \medskip
 
In another line of research, max-leaf problems have been studied 
in a purely combinatorial manner. For instance, for the undirected version, a well-known (tight) bound states that undirected graphs with minimum degree 3 have a spanning tree with at least $n/4+2$ leaves~\cite{KW91}. Similar bounds appear in~\cite{bonsmazickfeld,DingJS01}. 
 For digraphs, it is much harder to obtain tight bounds, or even bounds that are tight up to a constant factor.
Alon et al~\cite{AlonFGKS07II} showed that for strongly connected digraphs $D$ with minimum in-degree 3, $\lob(D)\geq \sqrt[3]{n/4}-1$ (this improves their previous bound from~\cite{AlonFGKS07}). In addition they construct 
strongly connected digraphs $D$ with minimum in-degree 3 with $\lob(D)=O(\sqrt{n})$. Considering the gap between this lower bound and upper bound, it is asked in~\cite{AlonFGKS07II} 
what the minimum value of $r$ is such that 
$\lob(D)\geq  f(n)\in \Theta(\sqrt[r]{n})$ for all graphs in this class ($2\leq r\leq 3$).

\medskip
{\em In this paper we answer this question by showing that for strongly connected digraphs $D$ with minimum in-degree 3, $\lob(D)\geq \frac{1}{4}\sqrt{n}$}. Considering the examples from~\cite{AlonFGKS07II}, we see that this bound is tight (up to a constant factor). 
Furthermore we generalize this result by showing that $\lob(D)\geq  f(n)\in  \Theta(\sqrt{n})$ holds for the larger class of digraphs with minimum in-degree 3 {\em without useless arcs}. {\em Useless arcs} are arcs that are not part of any out-branching.

\paragraph{Overview of new techniques}
The analysis that is required to prove the correctness of our improved algorithms is entirely new. 
However, the algorithms themselves are similar to those from~\cite{AlonFGKS07II}, and in particular to the one given in~\cite{ICALP08hopefully}. 
Therefore we give a short overview of the techniques used to obtain the previous FPT algorithms for \klot\ and \klob, and then give an overview of the new ideas and techniques in this paper.

In~\cite{ICALP08hopefully}, it is first observed that 
useless arcs may be deleted from the digraph.
For the resulting
digraph $D$, in~\cite{ICALP08hopefully} a variant of the algorithms introduced in~\cite{AlonFGKS07} and~\cite{AlonFGKS07II} is used: starting with an arbitrary out-branching, small changes are made that increase the number of leaves, until a locally optimal out-branching $T$ is obtained. 
{\em Back arcs} of $T$ are those arcs of $D$ that form a directed cycle together with a part of $T$. If at every point in $T$ (we omit the precise definition used in~\cite{ICALP08hopefully}) there are at most $6k^2$ back arcs, then a path decomposition of $D$ is constructed with width $w\leq 6k^3$, which allows for a dynamic programming procedure with complexity $2^{O(w\log w)}\cdot n$ to be used. On the other hand, if the number of back arcs is at least $6k^2$ at some point, it is shown that an out-branching with at least $k$ leaves exists. This last proof makes heavy use of the fact that no useless arcs are present.

In this paper, we construct a tree decomposition instead of a path decomposition (the locally optimal out-branching that we start with actually serves as the skeleton for the tree decomposition), and use a better way to group back arcs.
These two simple improvements do not only make the algorithm conceptually simpler, but also allow us to decrease the parameter function $2^{O(k \log^2 k)}$ for \klot\ from~\cite{AlonFGKS07II} by a logarithmical factor in the exponent to  $2^{O(k \log k)}$.

\begin{figure}[h]
\centering
\scalebox{1}{$\input{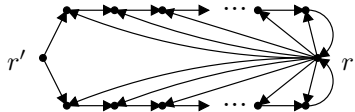}$}
\caption{A digraph without useless arcs with $\lot(D)=n-2$ (use $r$ as root) and $\lob(D)=(n-2)/2$ ($r'$ has to be the root).}
\label{fig:ratio}
\end{figure}

Our main technical contribution of this paper consists of two combinatorial bounds. The first of these bounds allows us to obtain a parameter function of $2^{O(k \log k)}$ also for \klob.
For out-branchings, our research is motivated by the following question: for digraphs without useless arcs, what is the highest possible ratio $\lot(D)/\lob(D)$?
Figure~\ref{fig:ratio} shows an example of a digraph without useless arcs where $\lot(D)/\lob(D)=2$.
In the first of the two main bounds of this paper, we prove that this ratio cannot be much larger; we prove that if $D$ contains no useless arcs, then $\lot(D)/\lob(D)\leq 3$. Since this ratio is bounded by a constant, an algorithm for \klob\ with the same complexity is then easily obtained.

To prove our second bound, the lower bound on $\lob(D)$ in strongly connected digraphs with minimum in-degree 3, we start with the method introduced in~\cite{AlonFGKS07II}: in~\cite{AlonFGKS07II} a locally optimal out-branching $T$ is considered. It is shown that if $T$ contains a path of length at least $2k^2$ that contains only vertices that have out-degree 1 in $T$, an out-branching with at least $k$ leaves can be found. If such a path does not exist, and $T$ itself also has less than $k$ leaves, the upper bound $n\leq 4k^3$ follows. In this paper we use the same general idea, but using a more sophisticated method to construct out-branchings, we can already find an out-branching with at least $k$ leaves if the aforementioned path has length $8k$.

The paper is organized as follows. Definitions and preliminary observations are given in Section~\ref{sect:prelim}. In Section~\ref{sec:FPT_DKL} the FPT algorithm for \klot\ is given, and in Section~\ref{sec:FPT_DSKL} the FPT algorithm for \klob\ is given. Section~\ref{sec:FPT_DSKL} also contains the proof that $\lot(D)/\lob(D)\leq 3$ for digraphs without useless arcs. In Section~\ref{sect:bound} we prove the lower bound for $\lob(D)$.

\section{Preliminaries}
\label{sect:prelim}

\paragraph{General definitions}

For basic graph theoretic definitions see~\cite{Diestel}, and for directed graphs in particular see~\cite{BangJG00}. 
We reuse many of the definitions and observations from~\cite{ICALP08hopefully} in this paper, so parts of this preliminaries section are taken literally from~\cite{ICALP08hopefully}.
For a digraph $D$, $V(D)$ denotes the set of vertices and $A(D)$ the set of arcs. Arcs are 2-tuples $(u,v)$ where $u\in V(D)$ is called the {\em tail} and $v\in V(D)$ the {\em head}.
For an arc set $B$, $\Hd(B)$ is the set of heads of arcs in $B$. 
A digraph $D$ is an {\em oriented graph} if $(u,v)\in A(D)$ implies $(v,u)\not\in A(D)$.
A {\em dipath} in a digraph $D$ is a sequence of distinct vertices $v_1,v_2,\ldots,v_r$ such that $(v_i,v_{i+1})\in A(D)$ for all $1\leq i\leq r-1$. This will also be called a {\em $(v_1,v_r)$-dipath}. The digraph consisting of these vertices and arcs will also be called a dipath.
With such a dipath we associate an order from $v_1$ to $v_r$, for instance when talking about the first arc of the path that satisfies some property.

A {\em partial order} is a binary relation that is reflexive, antisymmetric and transitive. A {\em strict partial order} is irreflexive and transitive. Partial orders will be denoted by $\preceq$, and strict partial orders by $\prec$.

For digraphs we will use normal (undirected) tree decompositions. Hence we define
a \emph{tree decomposition} of a digraph $D$ as a pair
$(X,U)$ where $U$ is an (undirected) tree whose vertices we will call \emph{nodes}, and 
$X=(\{X_{i} : i\in V(U)\})$ is a collection of subsets
of $V(D)$ (\emph{bags}) such that
\begin{enumerate}
\item
$\bigcup_{i \in V(U)} X_{i}= V(D)$, 
\item
for each arc $(v,w) \in A(D)$, there exists an $i\in
V(U)$ such that $v,w\in X_{i}$, and
\item
for each $v\in V(D)$, the set
of nodes $\{ i : v \in X_{i} \}$ forms a subtree of $U$.
\end{enumerate}
The \emph{width} of a tree decomposition $(\{ X_{i} : i \in
V(U) \}, U)$ equals $\max_{i \in V(U)} \{|X_{i}| - 1\}$. 
For notational convenience, we will also allow the graph $U$ in a tree decomposition $(X,U)$ to be directed, in this case it should be understood that we actually consider the underlying undirected graph of $U$.

\paragraph{Definitions for out-trees and out-branchings}

A subtree $T$ of a digraph $D$ is an \emph{out-tree} if it has only one vertex of in-degree zero, its \emph{root}. If $T$ is a spanning out-tree of $D$, i.e. $V(T)=V(D)$, then we call $T$  an \emph{out-branching} of $D$. 
The vertices of $T$ of out-degree zero are \emph{leaves} and the vertices of out-degree at least two are called \emph{branch vertices}.
Let $\LL(T)$ denote the set of leaves of $T$, let $\BR(T)$ denote the set of branch vertices of $T$, and let
$\BO(T)$ be the vertices of $T$ that have a branch vertex of $T$ as in-neighbor. 
Note that $\LL(T) \cap \BO(T)$ may not be empty.
\begin{propo}
\label{propo:ub_BO}
Let $T$ be an out-tree. Then $|\BO(T)|\leq 2|\LL(T)|-2$, and $|\BR(T)|\leq |\LL(T)|-1$.
\end{propo}
The omitted proofs in this section are straightforward and/or can be found in~\cite{AlonFGKS07II,ICALP08hopefully}.
If there exists a dipath in $D$ from vertex $u$ to vertex $v$, we say $v$ is {\em reachable} from $u$ (within $D$). The set of all vertices that are reachable from $u$ within $D$ is denoted by $R_D(u)$. (This set includes $u$ itself.)
\begin{propo}
\label{propo:extendingouttree}
Let $T$ be an out-tree of a digraph $D$, with root $r$. Then $D$ has an out-branching $T'$ with root $r$, that contains $T$, if and only if $R_D(r)=V(D)$.
\end{propo}
Let $T$ be an out-tree. Then we write $u\preceq_T v$ if $v\in R_T(u)$, and $u\prec_T v$ if in addition $v\not=u$. 
The following important observation will be used implicitly throughout the paper.
\begin{propo}
Let $T$ be an out-tree. The relation $\preceq_T$ is a partial order on $V(T)$.
\end{propo}

A digraph $H$ is {\em strongly connected} if for all pairs $u,v\in V(H)$, a $(u,v)$-dipath exists. A {\em strong component} is a maximal strongly connected subgraph. A strong component $H$ of $D$ is an {\em initial strong component} if there is no arc $(u,v)\in A(D)$ with $u\not\in V(H)$, $v\in V(H)$. 
Note that all initial strong components can be found in polynomial time.

Let $T$ be an out-branching of $D$, and let $(u,v)\in A(D)\bs A(T)$, where $v$ is not the root of $T$. The {\em 1-change for $(u,v)$} is the operation that yields $T+(u,v)-(w,v)$, where $w$ is the unique in-neighbor of $v$ in $T$.
We call an out-branching $T$ \emph{1-optimal} if there is no 1-change for an arc of $A(D) \bs A(T)$ that results in an out-branching $T'$ with more leaves. Note that a 1-optimal out-branching can be found in polynomial time.

\begin{propo}
\label{propo:onechange_allowed}
Let $T$ be an out-branching of $D$, and let $(u,v)\in A(D)\bs A(T)$. The 1-change for $(u,v)$ gives again an out-branching of $D$ if and only if $v\not\preceq_T u$.
\end{propo}

\begin{propo}
\label{propo:onechange_impr}
Let $T$ be an out-branching of $D$, and let $(u,v)\in A(D)\bs A(T)$. The 1-change for $(u,v)$ increases the number of leaves if and only if $u\not\in \LL(T)$ and $v\not\in \BO(T)$.
\end{propo}

An arc $(u,v)$ of a digraph $D$ is {\em useless} if $D$ has no out-branching containing $(u,v)$.
\begin{propo}
\label{propo:char_useless}
Let $D$ be a digraph with a vertex $r$ such that $R_D(r)=V(D)$. An arc $(u,v)$ of $D$ with $R_D(v)\not=V(D)$ is not useless if and only if there is a dipath in $D$ starting at $r$ that ends with $(u,v)$.
\end{propo}

Note that useless arcs can be removed in quadratic time.

\section{A Faster FPT Algorithm for \klot}
\label{sec:FPT_DKL}

We now show how back arcs of an out-tree are grouped, that is, how back arcs are assigned to vertices of the out-tree. 
Let $T$ be an out-tree of $D$ with $z\in V(T)$. Then
\[
\Back_D^T(z)=\{(u,v)\in A(D): v\prec_T z\preceq_T u\}.
\]
If it is clear what the graphs $D$ and $T$ in question are, the subscript and superscript will be omitted. When $|\Hd(\Back(z))|\geq k$ for some choice of $z$, an out-tree with at least $k$ leaves is easily found.

\begin{propo}
\label{propo:manybackarcs_outtree}
Let $T$ be an out-tree of $D$ with $|\Hd(\Back_D^T(z))|\geq k$ for some $z\in V(T)$. Then $D$ has an out-tree with at least $k$ leaves.
\end{propo}
\proof
Start with the out-tree $T[R_T(z)]$, which is rooted at $z$. For every vertex in $v\in \Hd(\Back_D^T(z))$, add an arc from some vertex in $u\in R_T(z)$ to $v$ (such an arc exists), making $v$ a leaf.\QED

This yields the correctness of Step~\ref{manybackarcs} of the algorithm, which is shown in Algorithm~\ref{alg:FPT_DKL}.

\begin{algorithm}[h]
    \dontprintsemicolon
    \Input{A digraph $D$ and integer $k$.}
    \BlankLine\;
      
    \For{every initial strong component $C$ of $D$ }{
        \lnl{pickroot}
	Choose $r\in V(C)$, let $D'=D[R_D(r)]$.\;
        \lnl{oneopt}
        Compute a 1-optimal out-branching $T$ of $D'$ with root $r$.\;	
        \lnl{enoughleaves}
	\lIf{$|\LL(T)|\geq k$}{Return(YES).}\;
        \lnl{manybackarcs}
	\lIf{there exists a vertex $z$ with $|\Hd(\Back_{D'}^T(z))|\geq k$}
	   {\\ \quad Return(YES).}\;
	\lnl{pathdecomp}
	Construct a tree decomposition of $D'$ with width at most $4k-5$.\;
        \lnl{dynprog}
        Do dynamic programming on the tree decomposition of $D'$.\;
        \lnl{outtreefound}
        \lIf{an out-tree with at least $k$ leaves is found}{Return(YES).}
    }
    \lnl{returnno}
    Return(NO)\;
    \caption{An FPT algorithm for \klot.}
    \label{alg:FPT_DKL}
\end{algorithm}

The construction of the tree decomposition of $D'$ is as follows. For the tree of the tree decomposition, we simply use the 
1-optimal out-branching $T$ itself. For a vertex $v\in V(T)$ with $(u,v)\in A(T)$, the bag $X_v$ of the tree decomposition is defined as follows.
\[
X_v =  \{u,v\}\ \cup\ \BO(T)\ \cup\ \LL(T)\ \cup\ \Hd(\Back_{D'}^T(v)).
\]
(If $v$ is the root of $T$, simply omit $u$.)
The tree decomposition is now $(X,T)$, with $X=\{X_v:v\in V(T)\}$.
\begin{lemma}
\label{lem:istreedecomp}
If $T$ is a 1-optimal out-branching of $D'$, then $(X,T)$ as constructed above is a tree decomposition of $D'$.
\end{lemma}
\proof
Every vertex $v\in V(D')$ is included in at least one bag, namely $X_v$. 
Now we show that for every arc $(u,v)\in A(D')$ there is a bag containing both $u$ and $v$. If one of its end vertices, say $v$, is in $\BO(T)$ or in $\LL(T)$, then $u,v\in X_u$. If $(u,v)\in A(T)$, then $u,v\in X_v$. Otherwise, since $T$ is 1-optimal, we have w.l.o.g. $v\prec_T u$ (Proposition~\ref{propo:onechange_allowed},~\ref{propo:onechange_impr}), and then we have $v\in \Hd(\Back(u))$, so $u,v\in X_u$. 

We now verify the third condition, namely that the vertex set $\mathcal{B}_v=\{u:v\in X_u\}$ induces a connected subgraph of $T$, for every $v\in V(T)$. If $v\in \LL(T)$ or $v\in \BO(T)$, then $\mathcal{B}_v=V(T)$, so the property obviously holds.
So now assume $v\not\in \LL(T)\cup \BO(T)$.
Suppose $v\in X_u$ for some $u\not=v$, so $v\in \Hd(\Back(u))$ or $(v,u)\in A(T)$. It then follows by the definition of $\Back(w)$
and the transitivity of $\preceq_T$
that for every $w$ with $v\preceq_T w\preceq_T u$, $v\in X_w$ holds. So $T[\mathcal{B}_v]$ contains a path from $v$ to $u$. This holds for every $u$ with $v\in X_u$, so this subgraph of $T$ is connected.
\QED

\begin{propo}
\label{propo:boundedwidth}
Let $T$ be an out-branching of a digraph $D$ 
with $|\LL(T)|\leq k-1$. If for all vertices $z\in V(D)$ it holds that $|\Hd(\Back_D^T(z))|\leq k-1$, then the tree decomposition $(X,T)$ as constructed above has width at most $4k-5$.
\end{propo}
\proof
This follows simply from 
\[
|X_u|=2+|\LL(T)|+|\BO(T)|+|\Hd(\Back(u))|\leq
\]
\[
2+ (k-1)+(2k-4)+(k-1)=4k-4,
\]
since $|\BO(T)|\leq 2|\LL(T)|-2$ (Proposition~\ref{propo:ub_BO}).\QED

When a tree decomposition is given of 
$D'$, standard dynamic programming methods can be used to decide whether $D'$ has an out-tree with at least $k$ leaves (see also~\cite{Bodl,Razgon08}).
The time complexity of such a procedure is $2^{O(w\log w)}\cdot n$, where $n=|V(D')|$ and $w$ is the width of the tree decomposition.

\begin{thm}
For any digraph $D$ with $n=|V(D)|$, Algorithm~\ref{alg:FPT_DKL} solves 
\klot\ in time $2^{O(k\log k)} \cdot n^{O(1)}$.
\end{thm}
\proof
Lemma~\ref{lem:istreedecomp} shows that the tuple $(X,T)$ we construct is indeed a tree decomposition. 

We now prove that Algorithm~\ref{alg:FPT_DKL} returns the correct answer in every case. Step~\ref{enoughleaves} and~\ref{outtreefound} are clearly correct. Step~\ref{manybackarcs} is correct by Proposition~\ref{propo:manybackarcs_outtree}.

To prove the correctness of Step~\ref{returnno}, suppose an out-tree $T$ with at least $k$ leaves exists in $D$. There is an initial strong component $C$ of $D$ such that $T$ is part of $D[R_D(r)]$ for any vertex $r\in V(C)$. In the iteration of the algorithm where $C$ is considered, the dynamic programming procedure of the tree decomposition will therefore return YES, if it is not returned before in Step~\ref{enoughleaves} or~\ref{manybackarcs}. So Step~\ref{returnno} only returns NO when no out-tree with at least $k$ leaves exists.

Finally we consider the time complexity of Algorithm~\ref{alg:FPT_DKL}. It is easy to see that every step of the algorithm can be done in time polynomial in $n$, except Step~\ref{dynprog}, which takes time $2^{O(k \log k)} \cdot n$, since the width of the tree decomposition is at most $4k-5$.
(Proposition~\ref{propo:boundedwidth}).
Steps~\ref{pickroot}--\ref{outtreefound} are repeated at most $n$ times (for every possible choice of initial strong component), so in total the complexity becomes $2^{O(k\log k)} \cdot n^{O(1)}$.\QED

Note that Algorithm~\ref{alg:FPT_DKL} can be made into a constructive FPT algorithm.

\section{A Faster FPT Algorithm for \klob}
\label{sec:FPT_DSKL}

We can modify the previous algorithm in order to solve \klob, see Algorithm~\ref{alg:FPT_DSKL}. 

\begin{algorithm}[h]
    \dontprintsemicolon
    \Input{A digraph $D$ and integer $k$.}
    \BlankLine\;

    \lnl{nooutbranching}
    \lIf{$D$ has no out-branching}{Return(NO).}\;
    \lnl{removearcs2}
    Remove from $D$ all useless arcs to obtain $D'$.\;
    \lnl{oneopt2}
    Compute a 1-optimal out-branching $T$ of $D'$.\;	
    \lnl{enoughleaves2}
    \lIf{$|\LL(T)|\geq k$}{Return(YES).}\;
    \lnl{manybackarcs2}
    \lIf{$T$ has a vertex $z$ such that $|\Hd(\Back_{D'}^T(z))|\geq 3k$}
	   {\\ \quad Return(YES).}\;
    \lnl{treedecomp2}
    Construct a tree decomposition of $D'$ with width at most $6k-5$.\;
    \lnl{dynprog2}
    Do dynamic programming on the tree decomposition of $D'$.\;
    \lnl{outbranchingfound2}
    \lIf{an out-branching with at least $k$ leaves is found}{Return(YES).}\;
    \lnl{returnno2}
    Return(NO)\;
    \caption{An FPT algorithm for \klob.}
    \label{alg:FPT_DSKL}
\end{algorithm}

The main new bound that we use to prove the correctness of algorithm~\ref{alg:FPT_DSKL} is proved later in Section~\ref{sec:mainthm}. There it is shown that if a digraph without useless arcs has an out-tree with at least $3k$ leaves, this can be used to construct an out-branching with at least $k$ leaves.
The tree decomposition used in the algorithm is exactly the same as the one constructed in Section~\ref{sec:FPT_DKL}. Since $|\Hd(\Back(z))|$ may now be at most $3k-1$, the width is at most $6k-5$.
\begin{propo}
\label{propo:boundedwidth2}
Let $T$ be a 1-optimal out-branching of a digraph $D$ 
with $|\LL(T)|\leq k-1$. If for all vertices $z\in V(D)$ it holds that $|\Hd(\Back_D^T(z))|\leq 3k-1$, then a tree decomposition $(X,T)$ of $D$ with width at most $6k-5$ can be constructed.
\end{propo}
We now prove the correctness of Algorithm~\ref{alg:FPT_DSKL}, and analyze its time complexity. 
\begin{thm}
For any digraph $D$ with $n=|V(D)|$, Algorithm~\ref{alg:FPT_DSKL} solves \klob\ in time $2^{O(k\log k)} \cdot n^{O(1)}$.
\end{thm}
\proof
We first prove that Algorithm~\ref{alg:FPT_DSKL} returns the correct answer in every case.
Step~\ref{nooutbranching},~\ref{enoughleaves2} and~\ref{outbranchingfound2} are obviously correct.
If $|\Hd(\Back(z))|\geq 3k$ for some $z$, an out-tree with at least $3k$ leaves exists (Proposition~\ref{propo:manybackarcs_outtree}), which in turn yields an out-branching with at least $k$ leaves since $D'$ contains no useless arcs (Theorem~\ref{thm:three_k}). This shows Step~\ref{manybackarcs2} is correct. 
If an out-branching of $D$ with at least $k$ leaves exists, then this is also an out-branching of $D'$, so in this case YES will be returned in Step~\ref{outbranchingfound2}, if not before. This proves the correctness of Step~\ref{returnno2}.

Finally we consider the time complexity of Algorithm~\ref{alg:FPT_DSKL}. 
Every step of the algorithm can be done in time polynomial in $n$, except Step~\ref{dynprog}, which takes time $2^{O(k \log k)} \cdot n$, since the width of the tree decomposition is bounded by $6k-5$ (Proposition~\ref{propo:boundedwidth2}). In total the complexity becomes $2^{O(k\log k)} \cdot n^{O(1)}$.\QED

\subsection{Constructing Leafy Out-Branchings from Out-trees}
\label{sec:mainthm}

In this section we prove one of the two main bounds of this paper, which yields the correctness of Step~\ref{manybackarcs2} of Algorithm~\ref{alg:FPT_DSKL}.
The proof of Theorem~\ref{thm:three_k} can be turned into a polynomial time algorithm that constructs an out-branching, and therefore Algorithm~\ref{alg:FPT_DSKL} can be made into a constructive FPT algorithm.

\begin{thm}
\label{thm:three_k}
Let $D$ be a digraph without useless arcs. If $\lot(D)\geq 3k$, then $\lob(D)\geq k$.
\end{thm}
\proof
Let $T$ be an out-tree of $D$ with at least $3k$ leaves, and let $r$ be the root of $T$. If $T$ contains at least one vertex $v$ with $R_D(v)=V(D)$, 
then also $R_D(r)=V(D)$, so then $T$ can be extended to an out-branching with at least $3k$ leaves (Proposition~\ref{propo:extendingouttree}).

\begin{figure}[h]
\centering
\scalebox{0.9}{$\input{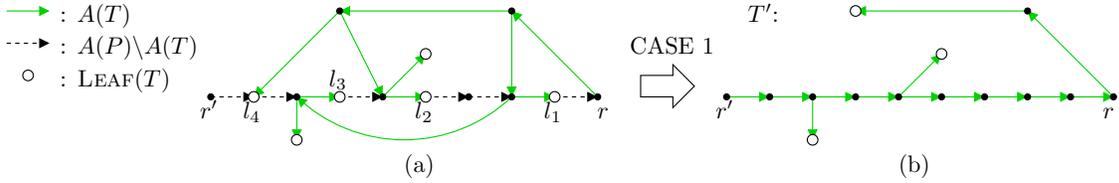}$}
\caption{(a) Out-tree $T$ and $(r',r)$-dipath $P$, and (b) the out-tree $T'$ constructed in Case 1.}
\label{fig:defs_case1}
\end{figure}

Otherwise, choose an arbitrary vertex $r'$ with $R_D(r')=V(D)$ (which exists since there are non-useless arcs, and thus at least one out-branching), and let $P$ be an $(r',r)$-dipath that contains a minimal number of vertices of $\LL(T)$. Let $\LL(T)\cap V(P)=\{l_1,\ldots,l_m\}$, labeled with decreasing labels along $P$. That is, if $i<j$, then $l_j\prec_P l_i$. These definitions are illustrated in Figure~\ref{fig:defs_case1}~(a).
We distinguish two types of vertices $l_i$ ($i\in\{1,\ldots,m\}$):
\begin{enumerate}
\item
Type 1: $D-l_i$ contains an $(x,y)$-dipath for some $x,y\in V(P)$ with $x\prec_P l_i\prec_P y$, with no internal vertices in $V(P)$.
\item
Type 2: all other vertices $l_i$.
\end{enumerate}
Now we consider three cases: since $|\LL(T)|\geq 3k$, one of the following holds: (i) $|\LL(T)\bs V(P)|\geq k$, (ii) the number of type 1 leaves is at least $k$, or (iii) the number of type 2 leaves is at least $k$. In all cases we will find an out-branching with at least $k$ leaves.\\
\\
CASE 1: $|\LL(T)\bs V(P)|\geq k$.\\
\\
We use $P$ and $T$ to construct an out-tree $T'$ of $D$. This is illustrated in Figure~\ref{fig:defs_case1}~(b). To construct $T'$, start with $T$. For all arcs $(u,v)\in A(P)$ with $v\not\in V(T)$ or $v=r$, simply add $(u,v)$ to the out-tree. For arcs $(u,v)\in A(P)\bs A(T)$ with $v\in V(T)\bs\{r\}$, do the 1-change for $(u,v)$. Then $T'$ is again an out-tree: for every vertex $v\in V(T')$, an $(r',v)$-dipath exists in $T'$,
every vertex except $r'$ has again in-degree 1, and $r'$ has in-degree 0. Also, for every vertex $v\in \LL(T)\bs V(P)$, the out-degree has not changed, so those vertices are still leaves. Thus we have an out-tree with at least $k$ leaves, with root $r'$ such that $R_D(r')=V(D)$. This is then easily extended to an out-branching with at least $k$ leaves (Proposition~\ref{propo:extendingouttree}).\\
\\
CASE 2: The number of type 1 leaves is at least $k$.\\
\\
\begin{figure}[h]
\centering
\scalebox{1}{$\input{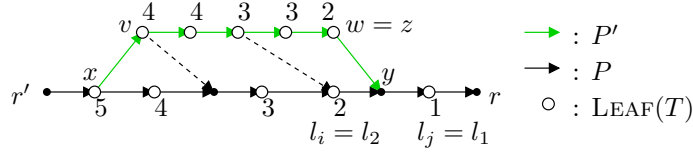}$}
\caption{Definitions used in Case 2. Numbers indicate $\LD$.}
\label{fig:defs_case2}
\end{figure}

\noindent
The definitions used in this case are illustrated in Figure~\ref{fig:defs_case2}.
For every $v\in \LL(T)$, we define the following value: if $r\in R_D(v)$, then consider the $(v,r)$-dipath of $D$ that contains the minimum number of $\LL(T)$-vertices. Then let $\LD(v)$ denote number of vertices in $\LL(T)$ on this path (including $v$ itself). 
Note that since we chose $P$ to contain the minimum number of $\LL(T)$-vertices, we have $\LD(l_i)=i$. In particular, all vertices $l_i$ receive different values for $\LD$.

\begin{figure}[h]
\centering
\scalebox{1}{$\input{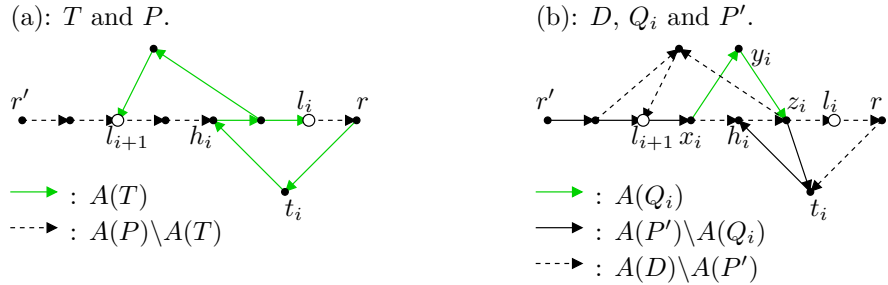}$}
\caption{Definitions used in Case 3.}
\label{fig:defs_case3}
\end{figure}

We now show that for every type 1 vertex $l_i$, there is a vertex $z\in \LL(T)\bs V(P)$ with $\LD(z)=\LD(l_i)$. Since $l_i$ is of type 1, we may consider an $(x,y)$-dipath $P'$ in $D$ with $x\prec_P l_i \prec_P y$ and no internal vertices in $P$. By choice of $P$, $P'$ contains at least one $\LL(T)$-vertex. Let $v$ be the first $\LL(T)$-vertex on $P'$, not equal to $x$. If $\LD(v)<\LD(l_i)$, then $P'$ can be used to find a path with fewer $\LL(T)$-vertices, a contradiction. So $P'$ contains an internal vertex $v$ with $\LD(v)\geq \LD(l_i)$. Now consider the maximum $j$ such that $y\preceq_P l_j$. By definition of $\LD$, $P'$ contains a $\LL(T)$-vertex $w\not=y$ with $\LD(w)\leq j+1\leq i$. So $P'$ also contains an internal vertex $w$ with $\LD(w)\leq \LD(l_i)$. Combining this with the fact that the $\LD$-labels decrease by steps of at most one when going along $P'$, it follows that $P'$ contains an internal vertex $z$ with $\LD(z)=\LD(l_i)$. Internal vertices of $P'$ are not part of $P$, so this proves that there is a vertex $z\in \LL(T)\bs V(P)$ with $\LD(z)=\LD(l_i)$, for every type 1 vertex $l_i$. Since we assumed there are at least $k$ type 1 vertices, and all of them receive different labels $\LD$, this proves that there are at least $k$ vertices in $\LL(T)\bs V(P)$, so by case 1 above, the desired out-branching exists.\\
\\
CASE 3: The number of type 2 leaves is at least $k$.\\
\\
In this case we will use the fact that $D$ contains no useless arcs. The definitions used are illustrated in Figure~\ref{fig:defs_case3}.

Let $l_i$ be a type 2 vertex. Consider the unique $(r,l_i)$-dipath in $T$. Let $(t_i,h_i)$ be the last arc of 
this path that is not in $A(P)$. Note that $h_i=l_i$ is possible. Note also that by choice of $(t_i,h_i)$, we have $l_{i+1}\prec_P h_i\preceq_P l_i$.
Since $(t_i,h_i)$ is not useless
and since we observed in the beginning of this proof that we may assume
$R_D(h_i)\not=V(D)$, there
is a dipath $P'$ in $D$ that starts in $r'$ and ends with the arc $(t_i,h_i)$ (Proposition~\ref{propo:char_useless}).
Let $x_i$ be the last vertex on $P'$ with $x_i\prec_P h_i$, and let $z_i$ be the first vertex on $P'$ after $x_i$ with $h_i\preceq_P z_i$. Since $r',h_i\in V(P')$, both vertices exist. Let $Q_i$ be the subpath of $P'$ from $x_i$ to $z_i$. So the internal vertices of $Q_i$ are not part of $P$, and $x_i\prec_P h_i\preceq_P z_i$. Combining this with $l_{i+1}\prec_P h_i\preceq_P l_i$ we obtain the following useful relations.
\[
l_{i+1}\prec_P z_i \hspace{2cm}
x_i\prec_P l_i
\]
Let $y_i$ be second vertex of $Q_i$. So $(x_i,y_i)\in A(Q_i)\bs A(P)$, though it is possible that $y_i\in V(P)$, 
namely when $y_i=z_i$.

Using these definitions, we can show how to construct an out-branching with at least $k$ leaves. Construct $T'$ as follows, starting with $P$. For every type 2 vertex $l_i$, if $y_i\not\in V(P)$, then add $(x_i,y_i)$. If $y_i\in V(P)$, then instead do the 1-change for $(x_i,y_i)$. 
In order to show that this yields again an out-tree, we need to prove that 
if $l_i$ and $l_j$ are two different type 2 vertices, then $y_i\not=y_j$. This is done below (Claim 1). Next we need to prove that for every type 2 vertex $l_i$, a leaf is gained. When $y_i\not\in V(P)$, this leaf will simply be $y_i$ itself. When $y_i\in V(P)$, then the corresponding leaf will be the in-neighbor $v$ of $y_i$ with respect to $P$ (so $(v,y_i)\in A(P)$). To prove that $v$ will indeed be a leaf, we need to show that the other operations do not increase its out-degree, hence that $v$ is not equal to $x_j$ for some other type 2 vertex $l_j$. This is also proved below (Claim 2).
Together this shows that $T'$ is an out-tree with root $r'$ with at least $k$ leaves, which is easily extended to the desired out-branching (Proposition~\ref{propo:extendingouttree}).\\

\noindent
{\bf Claim 1:} {\em For two type 2 vertices $l_i$ and $l_j$ with $i<j$,  $y_i\not=y_j$.}\\ 
Suppose $y_i=y_j$. Consider the path $Q_i$, and replace the first arc with the arc $(x_j,y_i)$. This gives an $(x_j,z_i)$-dipath with $x_j\prec_P l_j \preceq_P l_{i+1}\prec_P z_i$, which shows $l_j$ is in fact a type 1 vertex, a contradiction.\\

\noindent
{\bf Claim 2:} {\em If $y_i=z_i$ for some $i$ (so $(x_i,z_i)\in A(D)$), then 
there exists no type 2 vertex $j$ such that $(x_j,z_i)\in A(P)$.}\\
 To obtain a contradiction, assume that $(x_j,z_i)\in A(P)$. Note that $i\not=j$.
Since $l_{i+1}\prec_P z_i$ and $x_j\prec_P l_j$, it follows that $l_{i+1}\prec_P l_j$, so $i\geq j$. Using $i\not=j$ it follows that $i>j$, and therefore
\[
l_i\preceq_P l_{j+1}.
\]
We have an arc $(x_i,z_i)\in A(D)$ with $x_i\prec_P z_i$. By choice of $P$ it is not possible that $x_i\prec_P l_q\prec_P z_i$ for any $q$, since then a path containing fewer leaves of $T$ could have been chosen. But $x_i\prec_P l_i$, 
so also 
$z_i\preceq_P l_i$. Now we use the assumption that $(x_j,z_i)\in A(P)$, which yields $x_j\prec_P z_i\preceq_P l_i\preceq_P l_{j+1}\prec_P z_j$, so the path $Q_j$ shows that $l_i$ is in fact a type 1 vertex, a contradiction.\QED

\section{Lower bounds for the number of leaves}
\label{sect:bound}

The following lemma can be used for instance
to find leafy out-branchings in digraphs $D$ with minimum in-degree 3 (which is needed to satisfy the third condition). Its proof is postponed to the end of this section.

\begin{lemma}
\label{lem:outtreemainlem}
Let $T$ be 
an out-branching of a digraph $D$, and let $P=v_0,\ldots,v_{p-1}$ be a dipath in $T$ where 
\begin{itemize}
\item
$D$ contains no arcs $(v_i,v_j)$ with $i<j$, 
\item
$V(P)$ contains no branch vertices of $T$, and
\item
every $v_i$ has an in-neighbor in $D$ other than $v_{i-1}$ or $v_{i+1}$.
\end{itemize}
Then $D$ has an out-tree with at least $p/8$ leaves in $V(P)$.
\end{lemma}

Lemma~\ref{lem:outtreemainlem} is the key ingredient for our main result of this section. Apart from using this stronger lemma
and a shorter formulation, 
the proof of the next theorem is essentially the same as the one used in~\cite{AlonFGKS07II}.
\begin{thm}
\label{thm:outtreebound}
Let $D$ be a digraph on $n$ vertices with at least one out-branching. If $D$ has minimum in-degree 3, or if $D$ is an oriented graph with minimum in-degree 2, then $\lot(D)\geq \frac{1}{4}\sqrt{n}$.
\end{thm}
\proof
Let $k=\frac{1}{4}\sqrt{n}$. Consider a 1-optimal out-branching $T$ of $D$. We only have to consider the case that $|\LL(T)|\leq k-1$, and thus $|\BR(T)|<k-2$ (Proposition~\ref{propo:ub_BO}).
Consider the set $\mathcal{P}$ of all maximal dipaths in $T$ that contain no branch vertices. Note that every non-branch vertex of $T$ is in exactly one such path, so the paths in $\mathcal{P}$ give a partition of $V(T)\bs \BR(T)$. 
Note that every path in $\mathcal{P}$ either ends in a leaf of $T$, or ends in a vertex $u$ such that there is a branch vertex $v\in V(T)$ with $(u,v)\in A(T)$, and that for every branch vertex $v$ there is at most one such $u$. 
Hence the number of paths in $\mathcal{P}$ is bounded by $|\LL(T)|+|\BR(T)|\leq 2k-3$.

For every path $v_0,\ldots,v_{p-1}$ in $\mathcal{P}$ we may apply Lemma~\ref{lem:outtreemainlem}: since $D$ either has minimum in-degree 3 or is an oriented graph with minimum in-degree 2, every $v_i$ has an in-neighbor in $D$ other than $v_{i-1}$ or $v_{i+1}$. Since $T$ is 1-optimal, there are no arcs $(v_i,v_j)$ in $D$ with $i<j$ (Proposition~\ref{propo:onechange_allowed}, Proposition~\ref{propo:onechange_impr}). Hence if one of these paths contains at least $8k$ vertices, the desired out-tree exists
(Lemma~\ref{lem:outtreemainlem}). So finally suppose every path in $\mathcal{P}$ has less that $8k$ vertices. This yields
\[
n<8k(2k-3) + k-2< 16k^2,
\]
a contradiction with our choice of $k$. Hence in every case an out-tree with at least $\frac{1}{4}\sqrt{n}$ leaves can be found.\QED

Combining Theorem~\ref{thm:outtreebound} with Proposition~\ref{propo:extendingouttree} and Theorem~\ref{thm:three_k} respectively, we immediately obtain the following bounds for out-branchings.
\begin{cor}
\label{cor:mainlowerbound}
Let $D$ be a digraph on $n$ vertices that has minimum in-degree 3, or has minimum in-degree 2 and is an oriented graph. 
\begin{itemize}
\item
If $D$ is strongly connected, then $\lob(D)\geq\frac{1}{4}\sqrt{n}$.
\item
If $D$ contains no useless arcs, then $\lob(D)\geq \frac{1}{12}\sqrt{n}$.
\end{itemize}
\end{cor}

\noindent
It remains to prove Lemma~\ref{lem:outtreemainlem}. For this we will use the following lemma from~\cite{ICALP08hopefully}.

\begin{lemma}
\label{lem:changes_root_path}
Let $T$ be an out-branching of $D$ with root $r$.
Let $Q$ be a dipath in $D$ that starts at $r$. Then making all of the 1-changes for every arc in $A(Q)\bs A(T)$ yields again an out-branching of $D$ that contains $Q$.
\end{lemma}

\noindent
{\bf Proof of Lemma~\ref{lem:outtreemainlem}}:
Let $T$ be an out-branching of a digraph $D$, and let $P$ be a dipath in $T$ that satisfies the properties stated in the lemma.
Let $r$ be the root of $T$. 
If $v_{p-1}$ is not a leaf of $T$, then let $v_p$ be the unique out-neighbor of $v_{p-1}$ in $T$. In this case, we add the arc $(r,v_p)$ to $D$ (if it is not already present), and apply the 1-change for $(r,v_p)$ to $T$. So in both cases, from now on we may 
conveniently
assume that $R_T(v_i)=\{v_i,\ldots,v_{p-1}\}$. In the remainder of the proof we will use this to show that $D$ has an out-branching with at least $p/4$ leaves in $V(P)$. From this the statement follows; if we added $(r,v_p)$ then removing this arc from the out-branching will give two out-trees of the original digraph $D$, of which at least one has at least $p/8$ leaves in $V(P)$.

If an arc $(v_i,v_j)$ is present in $D$, then $i>j$. Arcs of this type are called {\em back arcs}. (Note that this is a subset of the arcs that were called back arcs in Section~\ref{sec:intro}.)

We will now iteratively make changes to $T$ until every $v_i\in V(P)$ is either a leaf, or is the tail of a back arc. The property of $T$ being an out-branching will be maintained throughout. To assist in the later analysis, tails of back arcs will be colored green or white as soon as these back arcs are added (details are given below).

Changes to $T$ are made in $p-1$ {\em stages}. During stage $i$ 
($i\in \{1,\ldots,p-1\}$), the goal is to make vertex $v_{i-1}$ a leaf, if 
this is still possible. For this we consider a dipath $Q_i$ that ends in the 
vertex $v_i$, and make 1-changes based on this path. The changes we make when 
considering the vertex $v_i$ will only involve arcs that are incident with 
vertices of $P$ with higher index, and vertices not in $P$. So in stages 
later than stage $i$, no changes are made to the arcs incident with $v_j$, 
for $j\leq i$. In particular, $v_{i-1}$ will remain a leaf if it is made a 
leaf in stage $i$.

\begin{figure}[h]
\centering
\scalebox{1}{$\input{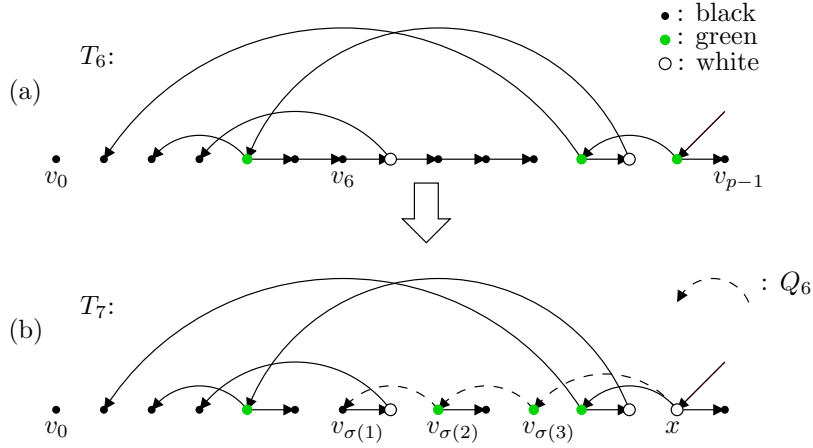}$}
\caption{Stage 6: constructing $T_7$ from $T_6$.}
\label{fig:boundproof1}
\end{figure}

Before we define $Q_i$, we observe that the following properties hold for $T$. These properties will be maintained throughout the procedure, and will therefore be called {\em invariant properties}. Note that the second property follows from the first.
\begin{enumerate}
\item
$v_i$ has only out-neighbors in $\{v_1,\ldots,v_{i+1}\}$, for all $i\in \{0,\ldots,p-1\}$.
\item
$R_T(v_i)\subseteq V(P)$.
\end{enumerate}
The changes that will be made to $T$ will consist of adding back arcs and adding arcs with tail not in $P$, and removing arcs of the form $(v_j,v_{j+1})$. Figure~\ref{fig:boundproof1}~(a) shows an example of how the out-branching may look after five stages (only the vertices of $P$ are shown). Note that the invariant still holds even though the set of reachable vertices may change for a vertex $v_i$.

The operation of stage $i$, and the dipath $Q_i$ that we use for it is defined as follows. Let $T_i$ denote the out-branching as it is in the beginning of stage $i$, so $T_1=T$.
The changes in stage $i$ will yield a new out-branching $T_{i+1}$.
In Figure~\ref{fig:boundproof1} an example is shown where $T_7$ is constructed from $T_6$. The dashed arcs in Figure~\ref{fig:boundproof1}~(b) show the dipath $Q_6$.
If $v_{i-1}$ is already a leaf or a tail of a back arc in $T_i$, we do nothing, so $T_{i+1}=T_i$. Otherwise, $v_i$ is the only out-neighbor of $v_{i-1}$ in $T_i$ (invariant Property~1). Then we consider a dipath $Q_i=x,v_{\sigma(q)},v_{\sigma(q-1)},\ldots,v_{\sigma(1)}$ in $D$ that ends in $v_i$, 
and has $x\not\in R_{T_i}(v_i)$, constructed as follows. Let $\sigma(1)=i$.
By our assumption, $v_i$ has an in-neighbor $u$ in $D$ that is not equal to $v_{i-1}$ or $v_{i+1}$. Since all arcs between vertices in $V(P)$ are back arcs and $R_{T_i}(v_i)\subseteq V(P)$ (invariant Property~2), this vertex $u$ is either not in $R_{T_i}(v_i)$ or it is equal to $v_j$ for some $j\geq i+2$. In the first case, $Q_i=u,v_i$. In the second 
case let $\sigma(2)=j$, and continue constructing the path using the same rule: $v_j$ has an in-neighbor that either is not in $R_{T_i}(v_j)$, or is equal to $v_l$ for some $l\geq j+2$, etc.

This process will terminate with a dipath $Q_i=x,v_{\sigma(q)},\ldots,v_{\sigma(1)}$, where $v_{\sigma(1)}=v_i$, the function $\sigma$ increases in steps of at least 2, and $x\not\in R_{T_i}(v_i)$. (Note that $x$ may or may not be in $V(P)$.)
It follows that if we make 1-changes for all arcs in $Q_i$, again an out-branching is obtained (Lemma~\ref{lem:changes_root_path}, 
note that we can easily extend $Q_i$ to start in $r$), and $v_{i-1}$ becomes a leaf. Observe also that the invariant properties are maintained by these changes. This yields $T_{i+1}$.

In addition we assign the following colors to vertices. All vertices of $T_1$ start out being black. In $T_{i+1}$ we color the vertices as follows:
\begin{itemize}
\item
The first vertex $x$ of $Q_i$ is colored white.
\item
The internal vertices $v_{\sigma(j)}$ of $Q_i$ ($j\in \{2,\ldots,q\}$) are colored green, unless they were already white in $T_i$.
\item
In all other cases vertices receive the same color as they have in $T_i$.
\end{itemize}
We say a {\em vertex becomes green (white) in stage $i$} if in $T_{i+1}$ it is green (white) but in $T_i$ it is colored differently.
We observe that the following {\em green vertex properties} hold:
\begin{enumerate}
\item
If $v_j$ is green in $T_i$, then $(v_{j-1},v_j)\not\in A(T_i)$.
\item
If a vertex $v_j$ becomes green in stage $i$, then $T_i$ contains the dipath $v_i,\ldots,v_j$.
\end{enumerate}

After $p-1$ stages, this procedure terminates with the out-branching $T_p$.
In order to give a lower bound for the number of leaves in $T_p$, we map all vertices that are tails of back arcs (green and white vertices) to leaves. This mapping is as follows:
\begin{itemize}
\item
A white vertex of $T_p$ that became white in stage $i$ is mapped to the leaf $v_{i-1}$ (note that this is still a leaf in $T_p$).
\item
A green vertex $v_j$ of $T_p$ is mapped to 
\begin{itemize}
\item
$v_{j-1}$ if it is a leaf,
\item
to the leaf $v_{i-1}$ if $v_{j-1}$ first became white in stage $i$, and
\item
to the leaf $v_{i-1}$ if $v_{j-1}$ is green and became green during stage $i$.
\end{itemize}
\end{itemize}
Finally, we show that every leaf has at most 3 preimages in this mapping. Consider a leaf $v_{i-1}$. If $v_{i-1}$ was already a leaf in $T_i$ (no changes are made in stage $i$), then only the 
vertex $v_i$ may be mapped to $v_{i-1}$, if $v_i$ is still green in $T_p$. 

On the other hand, if $v_{i-1}$ is made a leaf during stage $i$, then at the beginning of stage $i$, its 
unique 
out-neighbor is $v_i$, which therefore is not green (green vertex Property~1) and neither will be colored green during later stages. However, the single vertex $v_j$ that becomes white in stage $i$ is mapped to $v_{i-1}$. In addition, $v_{j+1}$ may be green in $T_p$, in which case this vertex is also mapped to $v_{i-1}$. Considering the above assignment rules, the only other vertices that may be mapped to $v_{i-1}$ are green vertices $v_j$ such that $v_{j-1}$ became green during stage $i$. We now argue that there is at most one such vertex. The vertex $v_j$ did not become green during a stage $s$ for $s>i$, since in stage $i$ the arc $(v_{j-2},v_{j-1})$ is removed, and $v_j$ only becomes green in stage $s$ if $v_s,v_{s+1},\ldots,v_j$ is a dipath in $T$ during stage $s$ (green vertex Property~2). On the other hand, if $v_j$ became green during an earlier stage, then during stage $i$ the arc $(v_{j-1},v_j)$ is not present anymore (green vertex Property~1). This means that $v_{j-1}$ must be the second vertex of the path $Q_i$,
since the in-neighbor of $v_{j-1}$ that is added to $Q_i$ is not in $R_{T_i}(v_i)$, so the construction of $Q_i$ ends after one more step. Hence there can only be one such vertex. 

This concludes the proof that every leaf has at most 3 preimages in our mapping. In addition, in $T_p$ every vertex of $V(P)$ is either a leaf, a green vertex, or a white vertex (note that $v_{p-1}$ starts out as a leaf). All white vertices and green vertices are mapped to leaves. It follows that at least $p/4$ vertices of $P$ end up being leaves. Together with the observation made in the beginning of the proof, the statement follows.\QED

\section{Discussion}

In Section~\ref{sec:mainthm} we showed that for digraphs $D$ without useless arcs, $\lot(D)/\lob(D)\leq 3$ holds. In Section~\ref{sec:intro} we gave a simple example where $\lot(D)/\lob(D)=2$. This leaves the question what the worst possible ratio may be. More complex examples exist where $\lot(D)/\lob(D)=2.5$.
Figure~\ref{fig:ratio_twopointfive} shows how to construct such a digraph $D$ consisting of $k$ digraphs on six vertices,
two extra vertices $r$ and $r'$, and arcs between these. 
This graph has an out-tree $T$ with $5k$ leaves (with root $r$), but it can be verified that $\lob(D)=2k$ (any out-branching needs to have $r'$ as root). 
For clarity, some arcs of $T$ are drawn as half arcs; these arcs should be understood as having $r$ as tail.
Observe that no arc in $D$ is useless.

\begin{figure}[h]
\centering
\scalebox{1}{$\input{ratio2point5.pstex_t}$}
\caption{A digraph $D$ with $\lot(D)/\lob(D)= 2.5$.}
\label{fig:ratio_twopointfive}
\end{figure}

We believe that this is the worst possible ratio. However bridging the gap between the factors 3 and $2.5$ may require a long proof and may not be worth the effort.

Similarly, we do not believe that the factor $\frac{1}{4}$ from Corollary~\ref{cor:mainlowerbound} is tight. We do not know what the best possible factor could be, but we do have examples showing that the analysis from Lemma~\ref{lem:outtreemainlem} is tight (up to a small additive term); to get a better factor, the construction of the out-branching
would need to be changed. Again we do not expect it to be worth the effort to improve the factors here.

It seems that in order to significantly improve the parameter function of FPT algorithms for these problems further, a different approach is needed, one that is not based on dynamic programming over a tree decomposition. 
Improving the factor from Theorem~\ref{thm:three_k} from 3 to 2.5 would for instance
slightly improve the constant that is suppressed by the $O$-notation in the expression $2^{O(k\log k)}$, but we do not consider this a significant improvement. It is an interesting question whether different, significantly faster FPT algorithms are possible for these two problems, 
for instance FPT algorithms with a parameter function of the form $c^k$ for some constant $c$. Such algorithms exist for the undirected version (with $c=6.75$, see~\cite{bonsmazickfeld}). This was also asked in~\cite{Razgon08}.

{\small

\bibliographystyle{siam}
\bibliography{digraph5}

\begin{thebibliography}{10}

\bibitem{AlonFGKS07II}
{\sc N.~Alon, F.~V. Fomin, G.~Gutin, M.~Krivelevich, and S.~Saurabh}, {\em
  Better algorithms and bounds for directed maximum leaf problems}, in FSTTCS
  07, vol.~4855 of LNCS, Springer, 2007, pp.~316--327.

\bibitem{AlonFGKS07}
\leavevmode\vrule height 2pt depth -1.6pt width 23pt, {\em Parameterized
  algorithms for directed maximum leaf problems}, in ICALP 07, vol.~4596 of
  LNCS, Springer, 2007, pp.~352--362.

\bibitem{BangJG00}
{\sc J.~Bang-Jensen and G.~Gutin}, {\em Digraphs: Theory, Algorithms and
  Applications}, Springer-Verlag, 2000.

\bibitem{Bodl}
{\sc H.~L. Bodlaender}, {\em A tourist guide through treewidth}, Acta
  Cybernet., 11 (1993), pp.~1--21.

\bibitem{BonsmaBW03}
{\sc P.~Bonsma, T.~Br{\"u}ggemann, and G.~J. Woeginger}, {\em A faster {FPT}
  algorithm for finding spanning trees with many leaves}, in MFCS 03, vol.~2747
  of LNCS, Springer, 2003, pp.~259--268.

\bibitem{ICALP08hopefully}
{\sc P.~Bonsma and F.~Dorn}, {\em An {FPT} algorithm for directed spanning
  k-leaf}.
\newblock http://arxiv.org/abs/0711.4052, submitted, 2007.

\bibitem{bonsmazickfeld}
{\sc P.~Bonsma and F.~Zickfeld}, {\em Spanning trees with many leaves in graphs
  without diamonds and blossoms}.
\newblock Accepted for LATIN 08. Preprint: http://arxiv.org/pdf/0707.2760,
  2007.

\bibitem{Diestel}
{\sc R.~Diestel}, {\em Graph Theory}, Springer-Verlag, 1997.

\bibitem{DingJS01}
{\sc G.~Ding, T.~Johnson, and P.~Seymour}, {\em Spanning trees with many
  leaves}, J. Graph Theory, 37 (2001), pp.~189--197.

\bibitem{DowneyF99}
{\sc R.~G. Downey and M.~R. Fellows}, {\em Parameterized complexity},
  Springer-Verlag, New York, 1999.

\bibitem{DrescherV07}
{\sc M.~Drescher and A.~Vetta}, {\em An approximation algorithm for the maximum
  leaf spanning arborescence problem}.
\newblock manuscript, 2007.

\bibitem{FellowsMRS00}
{\sc M.~R. Fellows, C.~McCartin, F.~A. Rosamond, and U.~Stege}, {\em
  Coordinatized kernels and catalytic reductions: An improved {FPT} algorithm
  for max leaf spanning tree and other problems}, in FSTTCS 00, vol.~1974 of
  LNCS, Springer, 2000, pp.~240--251.

\bibitem{FlumGrohebook}
{\sc J.~Flum and M.~Grohe}, {\em Parameterized Complexity Theory}, Texts in
  Theoretical Computer Science. An EATCS Series, Springer-Verlag, Berlin, 2006.

\bibitem{GK98}
{\sc S.~Guha and S.~Khuller}, {\em Approximation algorithms for connected
  dominating sets}, Algorithmica, 20 (1998), pp.~374--387.

\bibitem{Razgon08}
{\sc G.~Gutin, E.~Kim, and I.~Razgon}, {\em Minimum leaf out-branching
  problems}.
\newblock http://arxiv.org/abs/0801.1979v2, 2008.

\bibitem{KW91}
{\sc D.~J. Kleitman and D.~B. West}, {\em Spanning trees with many leaves},
  SIAM J. Discrete Math., 4 (1991), pp.~99--106.

\bibitem{Niedermeierbook06}
{\sc R.~Niedermeier}, {\em Invitation to fixed-parameter algorithms}, vol.~31
  of Oxford Lecture Series in Mathematics and its Applications, Oxford
  University Press, Oxford, 2006.

\bibitem{Solis-Oba98}
{\sc R.~Solis-Oba}, {\em 2-approximation algorithm for finding a spanning tree
  with maximum number of leaves}, in ESA 98, vol.~1461 of LNCS, Springer, 1998,
  pp.~441--452.

\bibitem{WuChao03}
{\sc B.~Wu and K.~Chao}, {\em Spanning Trees and optimization Problems}, CRC
  Press, 2003.

\end{thebibliography}
}

\end{document}